# Oriented Graphene Nanoribbons Embedded in Hexagonal Boron Nitride Trenches

*Lingxiu Chen,[1,2,†] Li He,[3,1,†] Hui Shan Wang,[4,1,†] Haomin Wang,[1,*] Shujie Tang,[1,5] Chunxiao Cong,[6,7]*

*Hong Xie,[1] Lei Li,[1,4] Hui Xia,[8] Tianxin Li,[8] Tianru Wu,[1] Daoli Zhang,[3] Lianwen Deng,[4] Ting Yu,[6]*

*Xiaoming Xie,[1,2,*] and Mianheng Jiang[1,2]*

*1 State Key Laboratory of Functional Materials for Informatics, Shanghai Institute of Microsystem and Information Technology, Chinese Academy of Sciences, 865 Changning Road, Shanghai 200050, P.R. China*
*2 School of Physical Science and Technology, ShanghaiTech University, 319 Yueyang Road, Shanghai 200031, P. R. China*
*3 School of Optical and Electronic Information, Huazhong University of Science and Technology, Wuhan, 430074, P. R. China*
*4 School of Physics and Electronics, Central South University, Changsha, 410083, P. R. China*
*5 Graduate University of the Chinese Academy of Sciences, Beijing 100049, P.R. China*
*6 Division of Physics and Applied Physics, School of Physical and Mathematical Sciences, Nanyang Technological University, 21 Nanyang Link, Singapore 637371*
*7 State Key Laboratory of ASIC & System, School of Information Science and Technology, Fudan University, Shanghai 200433, P.R. China*
*8 National Laboratory for Infrared Physics, Shanghai Institute of Technical Physics, Chinese Academy of Sciences, 500 Yu Tian Road, Shanghai 200083, P.R. China*
*† These authors contributed equally to this work*

*\* Electronic mail: hmwang@mail.sim.ac.cn, xmxie@mail.sim.ac.cn*

**Abstract:** Graphene nanoribbons (GNRs) are ultra-narrow strips of graphene that have the potential to be used in high-performance graphene-based semiconductor electronics. However, controlled growth of GNRs on dielectric substrates remains a challenge. Here, we report the successful growth of GNRs directly on hexagonal boron nitride substrates with smooth edges and controllable widths using chemical vapour deposition. The approach is based on a type of template growth that allows for the in-plane epitaxy of mono-layered GNRs in nano-trenches on hexagonal boron nitride with edges following a zigzag direction. The embedded GNR channels show excellent electronic properties, even at room temperature. Such in-plane hetero-integration of GNRs, which is compatible with integrated circuit processing, creates a gapped channel with a width of a few benzene rings, enabling the development of digital integrated circuitry based on GNRs.

## Introduction

Ideal graphene nanoribbons (GNRs) have been shown to exhibit extreme chirality dependence as metals or semiconductors.[1] Therefore, the capability to precisely produce GNRs with defined chirality at the atomic level is required in order to engineer their band gap and electrical properties.[2,3] Conventional lithography[4-6] always results in ragged edges along the GNRs. Other GNR synthesis methods, including sidewall growth on SiC,[7] advanced lithography,[8] sono-chemical methods[9,10] and carbon nanotube unzipping,[11,12] still present difficulties for either chirality control or width-scaling down to 10 nm and less. Recently, bottom-up synthesis methods using catalytic substrates[13-15] were demonstrated to form GNRs with well-defined edge structures and atomic precision. However, transferring techniques to produce devices without degrading the quality of the GNRs remain a formidable challenge. It is obvious that earlier approaches have fundamental limitations for further electronic investigation. Electronics always require scalable transfer-free approaches for growing GNRs and conducting band gap engineering. Controlled fabrication of oriented GNRs embedded on hexagonal boron nitride (h-BN) has the capability to overcome the above difficulties. With proper control, the band gap and magnetic properties can be precisely engineered. Most desired features for GNRs can be automatically attained using this approach.

Here, we demonstrate the successful growth of GNRs directly on hexagonal boron nitride (h-BN) substrates with smooth edges and controllable widths via templated growth using chemical vapour deposition (CVD). Transistors made with sub-10-nm GNRs demonstrate large on-off ratios of more than $10^4$ at room temperature and carrier mobility values of ~750 $cm^2V^{-1}s^{-1}$. For the narrowest GNRs, the band gaps extracted from the electrical transport data are greater than 0.4 eV.

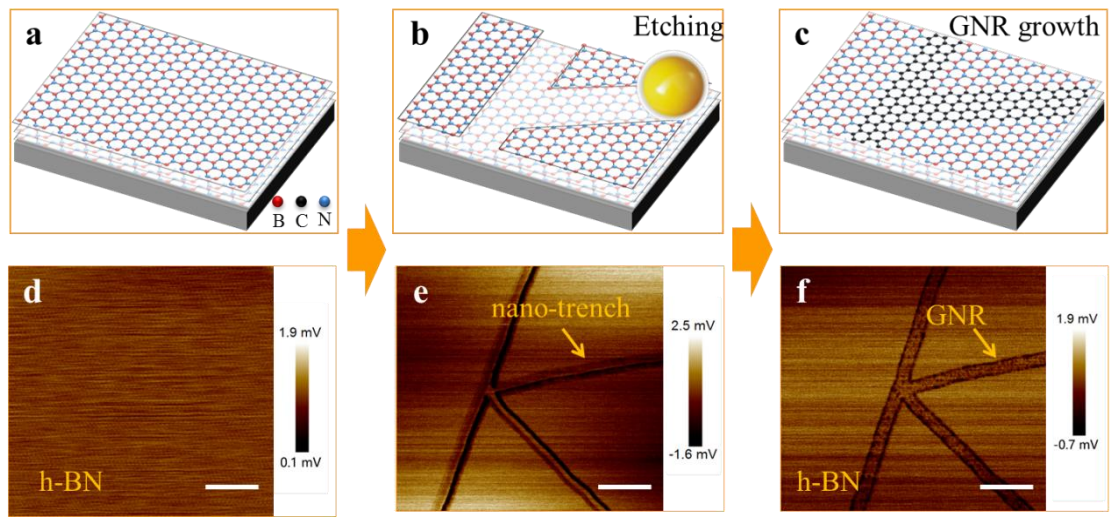

**Figure 1| Formation of graphene nanoribbons in h-BN trenches.** (a) Smooth surface of the h-BN; (b) Synthesis of nano-trenches on h-BN by Ni particle-assisted etching; (c) In-plane epitaxial template growth of graphene nanoribbons (GNRs) via chemical vapour deposition; (d-f) Atomic force microscopy (AFM) friction images corresponding to the schematics shown in (a-c), respectively. The friction images showed better contrast than the height images, especially for GNRs embedded in the h-BN nano-trenches. Scale bars: 200 nm.

## Results

### Templated growth of oriented GNRs in h-BN trenches

Fig. 1 demonstrates our conceptual design and its experimental verification for the synthesis of GNRs on h-BN via templated growth (details of the experimental procedures are given in the Methods). Single-crystal h-BN flakes exhibited a very smooth surface after multiple cleaning steps for removing possible surface contaminants (see the atomic force microscopy (AFM) images in Fig. 1d and Supplementary Fig. 1a and 1b). The nano-trenches on h-BN were synthesized by nickel particle-assisted etching (Fig. 1b and 1e and Supplementary Fig. 1c and 1d). After CVD growth, the trenches were filled with graphene (Fig. 1c and 1f and Supplementary Fig. 1e and 1f).

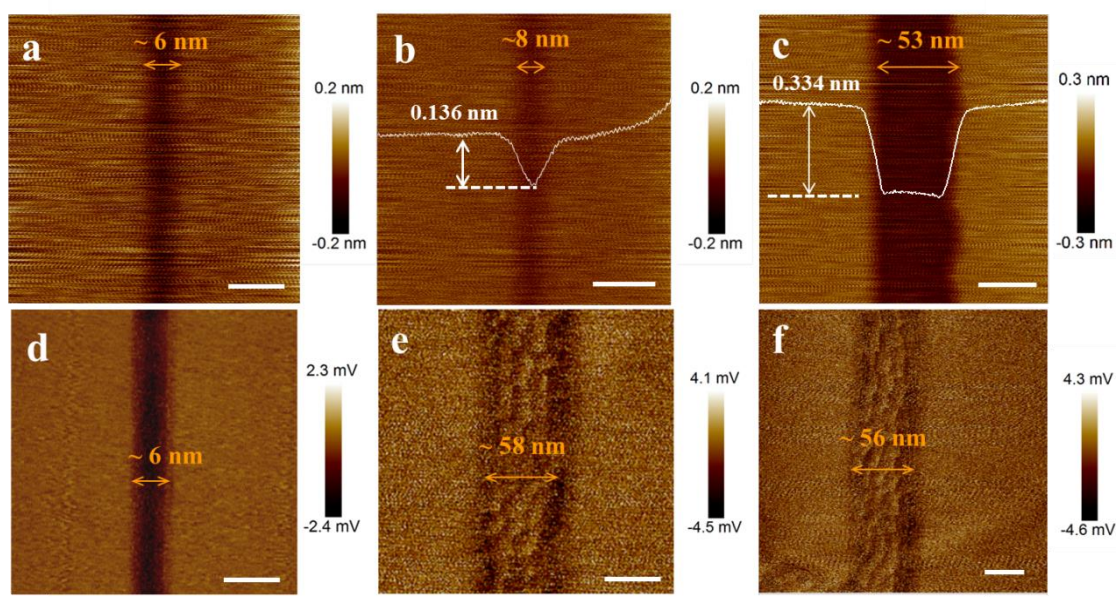

**Figure 2| Images of nano-trenches and graphene nanoribbons on the h-BN substrate.** AFM height images of monolayered nano-trenches on the h-BN crystal with the following widths: (a) ~6 nm, scale bar: 10 nm, (b) ~8 nm, scale bar: 20 nm and (c) ~53 nm, scale bar: 40 nm. AFM friction images of graphene nano-ribbons (GNRs) embedded in the nano-trenches on the h-BN crystal via template growth with the following widths: (d) ~6 nm, scale bar: 10 nm, (e) ~58 nm, scale bar: 40 nm and (f) ~56 nm, scale bar: 40 nm. The giant Moiré pattern can be observed in (e-f), indicating that the GNRs are precisely aligned with the h-BN.

### AFM measurement of oriented trenches in h-BN crystals

The nano-trenches exhibited predominant separation angles of 60° relative to each other with no obvious correlation to the direction of the gas flow, indicating anisotropic etching along the different crystallographic directions (Supplementary Figs. 2 and 3). Separation angles of 30° and 90° were rarely observed. The lattice structures shown in the inset of Supplementary Figs. 2a, 2b and 3a confirm that the crystallographic orientations of the h-BN trenches/edges follow a zigzag (ZZ) pattern. The crystallographically selective chemical reaction indicates lower activation energy along the zigzag patterns of the h-BN.

Fig. 2a-c shows some examples of nano-trenches obtained on h-BN substrates. The width of the trenches and their distribution were strongly dependent on the etching parameters and solutions (see Supplementary Tab. 1). Nanometre-sized-trenches, narrower than 10 nm in width, could be reproduced via an optimized process. Wider nano-trenches (Fig. 2c) could be synthesized by increasing the etching duration or temperature (see Supplementary Tab. 1). The trench in Fig. 2c had a depth of 0.334 nm with a bottom roughness comparable to that of pristine h-BN, indicating that the trench was monolayered. The extremely anisotropic two-dimensional nature of h-BN predominantly confined the etching to a single-atom layer. Due to the size limitation of the AFM tips, the depth of the narrower trenches could not be exactly determined, but these trenches were reasonably judged to also be mono-layered based on the measurements of wider nano-trenches. The tip size limitation may also result in uncertainty in the width measurements of the nano-trenches.

**AFM measurement of oriented GNRs on h-BN crystals**

Fig. 2d-f presents typical AFM friction images of GNRs embedded in the nano-trenches of the h-BN substrate, and the corresponding height images are given in Supplementary Fig. 4. A super-lattice structure, also known as a Moiré pattern, is clearly visible in Fig. 2e and 2f for the synthesized GNRs with widths of ~58 nm and ~56 nm, respectively, which was better contrasted in the friction images than in the height images. For narrow GNRs, a Moiré pattern was not observed (Fig. 2d and Supplementary Fig. 4a) because the GNR width was comparable to or smaller than the periodicity of the Moiré pattern. Height variations in the range of 20-40 pm could be seen near the boundary of the graphene/h-BN, which is most likely due to lattice distortions caused by mismatched lattice constants and coefficients of thermal expansion between the graphene and h-BN. The small out-of-plane distortions also excluded the possibility of the formation of multilayered GNRs. More height images of the GNRs are shown in Supplementary Fig. 5.

The existence of the giant Moiré pattern indicates that the graphene was highly crystalline and precisely aligned with the h-BN underneath. It was noticed that the Moiré pattern appeared to be stretched along the GNR, while it appeared relaxed laterally. This trend differs from regular hexagons with a periodicity of approximately 14 nm, which have always been observed with well-aligned graphene domains on h-BN.[16-18] This observation gives a strong indication of the in-plane epitaxy between the graphene and the h-BN at the edges of the trench, where the graphene is stretched by tensile strain along the ribbon, due to a lattice mismatch between the graphene and h-BN. Supplementary Fig. 6 illustrates how the stretched Moiré pattern formed, and Supplementary Fig. 7 shows the dependence of its wavelength along the stretched direction on the strain level. According to the relationship shown in Supplementary Fig. 7a, the strain in the GNR along the ribbon was estimated to be approximately 0.75%±0.3% from the wavelength of the stretched Moiré pattern measured in the AFM images. In addition, atomic-resolution AFM images confirmed that the in-plane connections between the GNRs and h-BN were continuous (see Supplementary Fig. 8).

Supplementary Fig. 9 shows AFM images taken before, during and after the GNR growth, illustrating that the graphene grew via a step-flow mechanism from two step-edges of the h-BN top-layer trench and coalesced into a complete GNR. The growth of the GNRs was also observed to occur at a one-sided atomic step-edge, which was formed by mechanical cleavage. Supplementary Fig. 10 shows a case in which the nanoribbon grew along the atomic step-edge of the top layer on the h-BN and developed laterally. Atomic-resolution images show that there were no discernible rotational misalignments between the three lattices (the graphene epi-ribbon, the top h-BN layer and the underlying h-BN lattice), indicating that while the GNR developed, the chirality of its edge was unchanged. Of course, in this case, the width of the GNR was determined by the growth time, not by the nano-trench confinement.

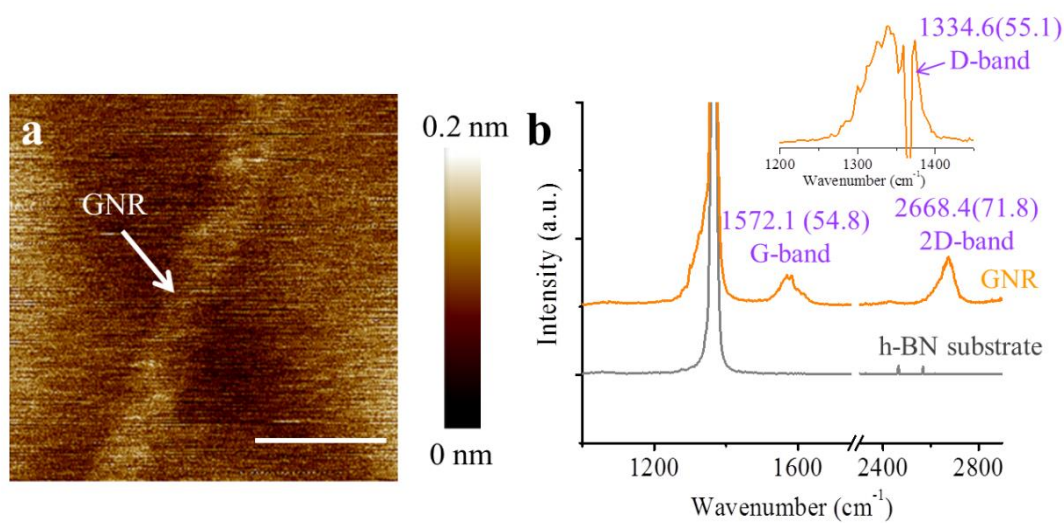

**Figure 3| Raman spectrum of a graphene nanoribbon on h-BN**. (a) AFM image of a graphene nanoribbon (GNR) with a width of ~15 nm. Scale bar: 30 nm; (b) Raman scattering of the GNR. The Raman spectrum of h-BN is also shown beside that of the GNR for comparison. The spectrum traces were normalized and shifted on the intensity axis for clarity. The inset shows the Raman spectrum of the GNR after subtracting the h-BN background. Because the position of the D-band of the GNR is very close to that of a prominent Raman peak of h-BN, such a subtraction could be utilized to identify the existence of weak Raman peaks. The full width at half-maximum (FWHM) for each peak is given in parentheses with the peak position. The band names of the main peaks are also indicated. The parameters of all Raman peaks were extracted via Lorentzian fitting, and the wavelength of the exciting laser was 488 nm.

**Raman characterization of GNRs embedded in the h-BN**

Raman measurements were carried out to investigate the structural and electronic properties of the nanoribbons. Fig. 3 shows the AFM image of a 15-nm-wide GNR and its Raman spectrum. In the spectra of Fig. 3b, a prominent sharp peak appeared at ~1365 cm$^{-1}$, which was attributed to the Raman-active LO phonon of h-BN[19]. For the spectrum of the GNR, the G-, D-, D'- and 2D-bands were fitted with a single Lorentzian line shape (see Supplementary Fig. 11). The Raman spectra show a prominent G-band (~1572.1 cm$^{-1}$) as well as a single Lorentzian-shaped 2D-band (~2668.4 cm$^{-1}$), which were expected for monolayer graphene. Compared to that of the pristine graphene domain[18], the red shift of ~9 cm$^{-1}$ in the G-band position indicates the existence of a tensile strain of approximately 0.6% in the GNR.[20,21] Furthermore, the GNR spectrum shows a D-band at 1334.6 cm$^{-1}$. At 1617.5 cm$^{-1}$ (D'-band), a tiny shoulder appears on the right side of the G band. Both the Raman D- and D'-bands may have primarily originated from the lattice distortions and disorder at the GNR/h-BN boundary. The 2D-band exhibits a characteristic of I(2D)/I(G) > 1, also indicating the single-layer nature of the GNR, where I(2D) and I(G) represent the intensity values of the 2D and G-bands, respectively. In addition, the observed broadening of the full width at half-maximum (FWHM) for the Raman G- and 2D-bands may have been due to strain variations at the nanometre-scale.[22]

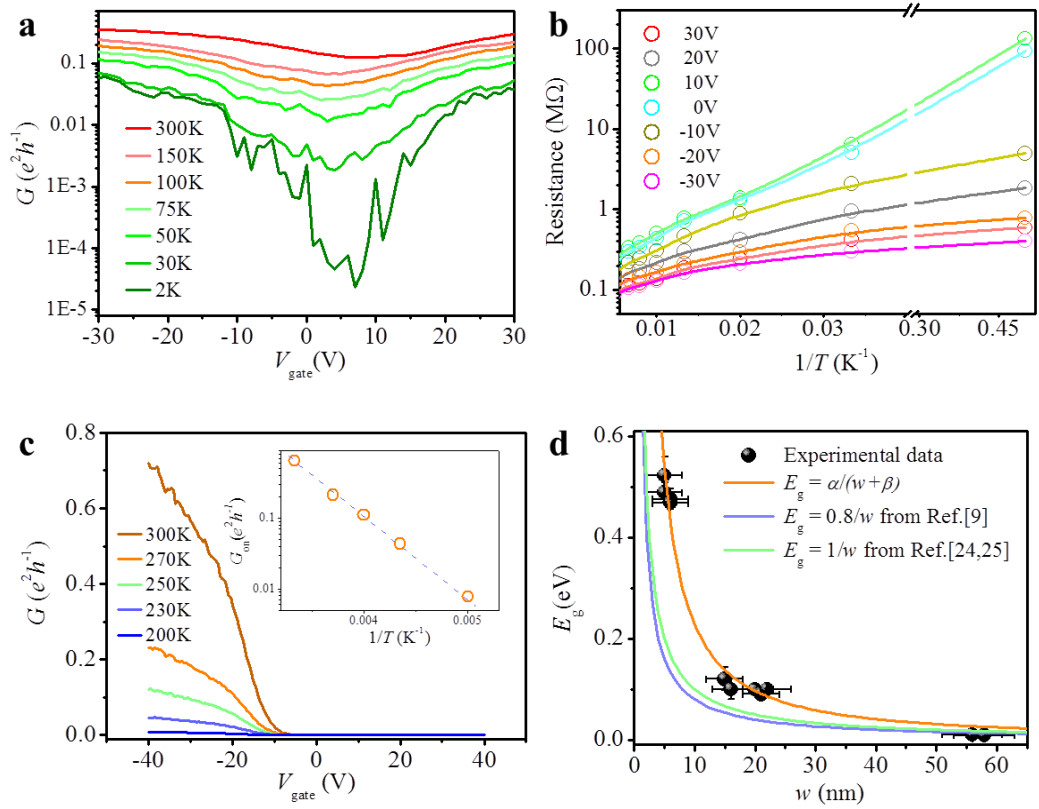

**Figure 4| Electronic transport through graphene nanoribbon devices on h-BN.** (a) Conductance (G) of graphene nanoribbons (GNRs) with a width of ~15 nm as a function of the back gate voltage ($V_{gate}$) at different temperatures; (b) Arrhenius plot of the resistance of the 15-nm GNR field effect transistor (FET) under different $V_{gate}$ values at temperatures from 2 to 250 K. The solid curves are fits based on a simple two-band (STB) model (see Supplementary Note 3); (c) Electronic transport through a narrow ribbon with a width of approximately 5 nm. Its conductance can be completely switched off, even at 300 K. The inset shows the conductance at $V_{gate}$=-30 V versus inverse temperature from 200 to 300 K. The dashed line is fit to the experimental data according to $G_{on} \sim \exp(\frac{E_g}{2k_BT})$; (d) Band gap $E_g$ extracted from experimental data for GNRs versus their ribbon width (w). The orange curve is a fit of our experimental data with $E_g(eV) \sim \alpha/(w+\beta)$ [α is in units of eV nm, both β and w are in units of nm]. The light purple and light green curves are empirical curves adapted from [ref. 9] and [refs. 24, 25], respectively. Error bars for the experimental data represent standard deviation of uncertainty in AFM measurement or gap extraction.

**Electronic transport properties of GNR transistors**

To investigate the electrical properties of the GNRs, field effect transistor (FET) devices were produced using p-type doped Si/SiO$_2$ (300 nm thick for the SiO$_2$ layer) (for details of the device fabrication, see Methods and Supplementary Figs. 11-16). Fig. 4 shows the field effect characteristics of representative GNRs at different widths. For the 15-nm GNR, the plot of conductance $G$ versus $V_{gate}$ at different temperatures, shown in Fig. 4a, exhibits obvious modulations due to the external electrical field. The field effect mobility extracted from the 15-nm GNR was ~916.1 cm$^2$V$^{-1}$s$^{-1}$ at 300 K. According to a simple two-band (STB) model (see Supplementary Note 3), the value of the band gap could be extracted by fitting the resistance-temperature curve (Fig. 4b) (for details of the fitting, please see Supplementary Information). The fitting includes contributions from both the normal thermal activation and electrical contact resistance. The extracted band gap for the 15-nm GNR was 120±23 meV. Next, the narrowest GNRs were measured. Some devices exhibited obvious transistor behaviour, even at room temperature. As shown in Fig. 4c, a ~5-nm GNR device had a $G_{on}/G_{off}$ ratio of >10$^4$. Its conductance appeared to be completely insulating, with $G<10^{-4}$ $e^2$ $h^{-1}$=10$^{-10}$ S from $V_{gate}$=0 to 10 V (off-state); then, it gradually switched on, exhibiting a relatively high $G$=0.7 $e^2$ $h^{-1}$ at room temperature for $V_{gate}$=-40 V. The scattering mean free path (MFP) was estimated to be approximately 50 nm. The mobility of the GNR was estimated to be approximately 765 cm$^2$V$^{-1}$s$^{-1}$ at 300 K. Accurately measuring the off-state resistance of the narrowest GNR FET became challenging due to noise and setup limitations. It was found that the conductance near the on-state exhibited an exponential relationship with inverse temperature from 200 to 300 K. This result indicates that a Schottky barrier (SB) dominated the conductance of the GNR FET because of the high work functions of Pd and Ni[9,23]. We estimated the band gaps ($E_g$) of the narrow GNRs by fitting the conductance near the on-state $G_{on} \propto \frac{I_{on}}{I_{off}} \sim \exp(\frac{E_g}{2k_B T})$, where $k_B$ is Boltzmann's constant and $T$ is temperature. The extracted band gap value for the 5-nm GNR was 489.4±19.0 meV (see the Supplementary Information for details of the fitting process). Using similar methods, the band gap values for other narrow GNRs were also estimated. As shown in Fig. 4d, the band gaps, $E_g$, extracted from all measured GNRs were plotted with respect to the width of the corresponding GNRs. It is obvious that the band gap scaled inversely with ribbon width. The width dependence of the band gap fit well with the function $E_g$ (eV) ~$\alpha/(w+\beta)$ [$w$ is in units of nm], where parameter $\alpha \approx$1.99 eV nm and parameter $\beta$=-1.28 nm.

**Discussion**

In the GNR samples, the band gap exhibited a strong dependence on the width of the ribbons. In particular, the sub-10-nm GNRs exhibited an electronic band gap of ~0.5 eV. It is important to understand the possible origins of the band gap. Normally, the electronic structure of GNRs is strongly dependent on the edge of the GNRs. For this study, it was challenging to characterize the exact geometry of the narrow ribbons obtained (for example, the edge chirality and termination state of the bonds). However, the GNRs were generally parallel to the ZZ pattern of the h-BN and free of detectable edge roughness. Therefore, it is likely that most segments of the GNR edges followed a ZZ pattern. Previous studies in the literature predicted that GNRs with pristine ZZ edges would always be metallic because of their peculiar flat-band edge states localized near the Fermi level.[1] Recent theoretical papers have predicted that the flat-band would split due to electron-electron interactions, followed by the opening of a band gap caused by antiferromagnetic coupling between the spins along opposite edges of the ZZ ribbons.[24,25] More recently, band gaps in narrow ZZ nanoribbons have been experimentally observed in scanning tunnelling microscopy (STM) and scanning tunnelling spectroscopy (STS) studies.[26,27] A similar high density of states (DOS) flat-band near the Fermi level was also observed in ZZ-terminated atomically sharp graphene−BN interfaces.[28] We believe that the gap opening in the GNRs embedded in h-BN may also have been due to *e-e* interactions. It is noted that our narrow GNRs with similar widths exhibited relatively larger band gaps than those reported in previous literature [refs. 9 and 24]. The uniaxial strain[29] from in-plane graphene-BN bonding and the Bernal stacking[30] on h-BN may have provided additional contributions to the gap opening in the GNRs embedded in h-BN.

The applicability of graphene for future digital devices is often questioned due to its intrinsic gapless nature. Nanoribbons offer a potential solution, but both the width and edges must be precisely controlled. In-plane graphene-BN hetero-structured films have been reported, however, only on metal surfaces[31-35]. In addition, control over the dimensions of graphene and graphene-BN boundaries has not been fully achieved. By employing the in-plane epitaxy of graphene in nano-trenches of h-BN, we have realized ZZ-oriented graphene nanoribbons with a controlled width and smooth edges. The GNRs feature a tunable band gap, enabling sub-10-nm GNR FETs with on-off ratios greater than $10^4$. Our results demonstrate that it is possible to resolve the fundamental gapless limitation of graphene, paving the way for the realization of graphene-based digital electronics that can operate at room temperature.

**Methods**

**Etching process on h-BN surface**
To prepare h-BN samples for nano-particle-assisted etching, multiple cleaning steps were adapted to ensure the reliable production of the cuts. Before depositing the h-BN flake, quartz substrates were cleaned with acetone and isopropyl alcohol and then annealed at 600 ℃ for 30 min to remove organic contaminants. Subsequently, the h-BN was mechanically exfoliated using semiconductor-grade tape. Next, the h-BN samples were heated in a quartz tube at 500 ℃ for 15 min under an $Ar:H_2$ flow (850:150 sccm) in order to remove the tape residue. After the heat cleaning process, a solution of $NiCl_2:H_2O$ at a concentration of 0.01 mg mL$^{-1}$ was spun at 1800 rpm for 100 s onto the substrate surface and then baked for 10 min at 80 ℃ on a hot plate to evaporate the solvent. This $NiCl_2$-treated sample was then submitted to a two-step process under an $Ar:H_2$ flow (850:150 sccm): annealing at 500 ℃ for 20 min, which resulted in Ni nano-particle formation, and etching at 1200 ℃ for 60-180 min. The optimal etching pressure was found to be approximately 150 Pa.

**Growth of GNRs**
Before graphene growth, etched h-BN flakes were cleaned separately in HCl solution, DI water and acetone. The substrate with the h-BN flakes was then loaded into a growth chamber. GNR growth was carried out in a low-pressure CVD furnace at 1280 ℃ under an Ar flow of 10 sccm, corresponding to 15 Pa; the samples were then annealed for 5 min. Next, the Ar flow was turned off, and a $C_2H_2$ flow and a mixture of silane and Ar (5% mole ratio of silane to Ar) were introduced into the system for GNR growth. The ratio of $C_2H_2$ to silane was approximately 1:1. The pressure was maintained at 5 Pa during the growth process, and the growth time was approximately 5 min. After growth, both the $C_2H_2$ and silane/Ar flow were turned off, and the system was cooled to room temperature with flowing Ar. The samples of GNRs on h-BN grown on a quartz surface were moved to a highly p-type doped silicon wafer with a 300-nm-thick $SiO_2$ capping layer for electrical transport studies.

**Atomic force microscopy**

Cleaned samples were characterized using one AFM (Dimension Icon, Bruker), while atomic-resolution images were taken by another AFM (Multimode IV, Veeco) under ambient conditions. AFM measurements were acquired in contact mode to obtain height and friction images. SNL-10 AFM tips from Bruker, which possess a nominal tip radius of less than 10 nm, were used in all measurements. The use of friction contrast was necessary because this mode gives clear information about the super-structure and atomic lattice. For atomic-resolution scanning, the force constant $k$ of the cantilever tips was in the range of 0.05-0.5 N m$^{-1}$. The scan rate was set to 10-60 Hz to minimize any noise from thermal drift. The integral gain and set point were adjusted to be as low as possible during the measurement. Several hours of pre-scanning were carried out to warm up the scanner to obtain good stability during imaging. To ensure a highly accurate atomic-resolution image, scanners with a travel range of less than 10 μm along the x and y directions were used. Calibration at atomic resolution was performed with newly cleaved highly ordered pyrolytic graphite (HOPG) before measurement.

**Characterization of GNRs using Raman spectroscopy**

Raman spectra were obtained with a commercially available confocal Raman instrument: model Alpha 300R from WITec. The Raman data were recorded using a laser wavelength of 488 nm ($\hbar\omega_L$=2.54 eV). An objective lens with 100x magnification and a 0.95 numerical aperture was used, producing a laser spot that was approximately 500 nm in diameter. The laser power was maintained at less than 1 mW on the sample surface to avoid laser-induced heating.

**Device fabrication and transport measurements**

GNR devices were created by a standard electron beam lithographic technique with Ni or Pd as the source and drain contacts on p-doped silicon wafers with 300-nm-thick $SiO_2$. Next, the devices were annealed in a hydrogen atmosphere at 200 °C for 3 h to remove the resist residues and to reduce the contact resistance between the GNRs and metal electrodes before electrical measurements because the thickness of the h-BN flakes on the silicon wafer was approximately 15 nm. Electrical transport measurements were carried out using a physical property measurement system (PPMS from Quantum Design, Inc.) via a Keithley 4200 semiconductor characterization system.

**Data availability**

The authors declare that the main data supporting the findings of this study are available within the article and its Supplementary Information files. Additional data are available from the corresponding author upon request.

**References**

[1] Nakada, K., Fujita, M., Dresselhaus, G. & Dresselhaus, M. S. Edge state in graphene ribbons: Nanometer size effect and edge shape dependence. *Physical Review. B, Condensed matter* **54**, 17954-17961 (1996).

[2] Son, Y. W., Cohen, M. L. & Louie, S. G. Half-metallic graphene nanoribbons. *Nature* **444**, 347-349 (2006).

[3] Novoselov, K. S. *et al.* A roadmap for graphene. *Nature* **490**, 192-200 (2012).

[4] Han, M. Y., Ozyilmaz, B., Zhang, Y. & Kim, P. Energy band gap engineering of graphene nanoribbons. *Phys Rev Lett* **98**, 206805 (2007).

[5] Chen, Z. H., Lin, Y. M., Rooks, M. J. & Avouris, P. Graphene nano-ribbon electronics. *Physica E* **40**, 228-232 (2007).

[6] Bai, J., Zhong, X., Jiang, S., Huang, Y. & Duan, X. Graphene nanomesh. *Nature nanotechnology* **5**, 190-194 (2010).

[7] Sprinkle, M. *et al.* Scalable templated growth of graphene nanoribbons on SiC. *Nature nanotechnology* **5**, 727-731 (2010).

[8] Tapaszto, L., Dobrik, G., Lambin, P. & Biro, L. P. Tailoring the atomic structure of graphene nanoribbons by scanning tunnelling microscope lithography. *Nature nanotechnology* **3**, 397-401 (2008).

[9] Li, X., Wang, X., Zhang, L., Lee, S. & Dai, H. Chemically derived, ultrasmooth graphene nanoribbon semiconductors. *Science* **319**, 1229-1232 (2008).

[10] Wu, Z. S. *et al.* Efficient Synthesis of Graphene Nanoribbons Sonochemically Cut from Graphene Sheets. *Nano Research* **3**, 16-22 (2010).

[11] Jiao, L., Zhang, L., Wang, X., Diankov, G. & Dai, H. Narrow graphene nanoribbons from carbon nanotubes. *Nature* **458**, 877-880 (2009).

[12] Kosynkin, D. V. *et al.* Longitudinal unzipping of carbon nanotubes to form graphene nanoribbons. *Nature* **458**, 872-876 (2009).

[13] Cai, J. *et al.* Atomically precise bottom-up fabrication of graphene nanoribbons. *Nature* **466**, 470-473 (2010).

[14] Jacobberger, Robert M. *et al.* Direct oriented growth of armchair graphene nanoribbons on germanium. *Nature communications* **6**, 8006 (2015).

[15] Ago, H. *et al.* Lattice-Oriented Catalytic Growth of Graphene Nanoribbons on Heteroepitaxial Nickel Films. *ACS Nano*, **7**, 10825-10833 (2013).

[16] Tang, S. *et al.* Precisely aligned graphene grown on hexagonal boron nitride by catalyst free chemical vapor deposition. Sci. Rep. **3**, 2666 (2013).

[17] Xue, J. *et al.* Scanning tunnelling microscopy and spectroscopy of ultra-flat graphene on hexagonal boron nitride. *Nature materials* **10**, 282-285 (2011).

[18] Tang, S. *et al.* Silane-catalysed fast growth of large single-crystalline graphene on hexagonal boron nitride. *Nature communications* **6**, 6499 (2015).

[19] Geick, R., Perry, C. H. & Rupprech.G. Normal Modes in Hexagonal Boron Nitride. *Physical Review* **146**, 543-547 (1966).

[20] Mohiuddin, T. M. G. *et al.* Uniaxial strain in graphene by Raman spectroscopy: G peak splitting, Grüneisen parameters, and sample orientation. *Phys. Rev. B* **79**, 205433 (2009).

[21] Huang, M. *et al.* Phonon softening and crystallographic orientation of strained graphene studied by Raman spectroscopy. *PNAS* **106**, 7304–7308 (2009).

[22] C. Neumann, *et al.* Raman spectroscopy as probe of nanometre-scale strain variations in graphene. *Nature communications* **6**, 8429 (2015).

[23] Javey, A., Guo, J., Wang, Q., Lundstrom, M. & Dai, H. Ballistic carbon nanotube field-effect transistors. *Nature* **424**, 654-657 (2003).

[24] Son, Y. W., Cohen, M. L. & Louie, S. G. Energy gaps in graphene nanoribbons. *Phys. Rev. Lett.* **97**, 216803 (2006).

[25] Yang, L., Park, C. H., Son, Y. W., Cohen, M. L. & Louie, S. G. Quasiparticle energies and band gaps in graphene nanoribbons. *Phys. Rev. Lett.* **99**, 186801 (2007).

[26] Magda, G. Z. *et al.* Room-temperature magnetic order on zigzag edges of narrow graphene nanoribbons. *Nature* **514**, 608–611 (2014).

[27] Li, Y., Chen, M., Weinert, M. & Li, L. Direct experimental determination of onset of electron–electron interactions in gap opening of zigzag graphene nanoribbons. *Nature communications* **5**, 4311(2014).

[28] Drost, R. *et al.* Electronic states at the graphene–hexagonal boron nitride zigzag interface. *Nano letters* **14**, 5128-5132 (2014).

[29] Pereira, V. M., Neto, A. H. C. and Peres, N. M. R. Tight-binding approach to uniaxial strain in graphene. *Phys. Rev. B* **80**, 045401 (2009).

[30] Hunt, B. *et al.* Massive Dirac fermions and Hofstadter butterfly in a van der Waals heterostructure. *Science* **340**, 1427-1430 (2013).

[31] Ci, L. *et al.* Atomic layers of hybridized boron nitride and graphene domains. *Nature materials* **9**, 430-435 (2010).

[32] Liu, L. *et al.* Heteroepitaxial growth of two-dimensional hexagonal boron nitride templated by graphene edges. *Science* **343**, 163-167 (2014).

[33] Levendorf, M. P. *et al.* Graphene and boron nitride lateral heterostructures for atomically thin circuitry. *Nature* **488**, 627-632 (2012).

[34] Liu, Z. *et al.* In-plane heterostructures of graphene and hexagonal boron nitride with controlled domain sizes. *Nature nanotechnology* **8**, 119-124 (2013).

[35] Gong, Y. J. *et al.* Direct chemical conversion of graphene to boron- and nitrogen- and carbon-containing atomic layers. *Nature communications* **5**, 3193 (2014).

**Acknowledgements**

The work performed at the Shanghai Institute of Microsystem and Information Technology, Chinese Academy of Sciences, was partially supported by the National Science and Technology Major Projects of China (grant no. 2011ZX02707), the Chinese Academy of Sciences (Grant Nos. KGZD-EW-303 and XDB04040300), a project from the Science and Technology Commission of Shanghai Municipality (Grant No. 16ZR1442700) and the National Science Foundation of China (Grant No. 61136005). The work conducted at the Shanghai Institute of Technical Physics, Chinese Academy of Sciences, was partially supported by the National Science Foundation of China (Grant No. 91321311). T.Y. is grateful for the support of MOE2012-T2-049. H.W. and X.X. thank T. Taniguchi, K. Watanabe (National Institute for Materials Science, Japan) and J. H. Edgar (Kansas State University, U.S.A.) for supplying the partial h-BN crystals.

**Author contributions**

H.W. and X.X. directed the research work. H.W. conceived and designed the experiments. L.H. performed the etching processes on h-BN. L.C. performed the growth experiments for the GNRs. S.T. carried out both the etching and GNR growth during the beginning stage of this project. L.H., L.X.C., H.XIA and T. L. performed the AFM experiments. H.S.W., H.W. and H.XIE fabricated the electronic devices. H.S.W. and L.L. performed the transport measurements. H.W. and L.X.C. performed the calculations concerning the stretched Moiré pattern. L.X.C., C.C. and T.Y. performed the Raman measurements. H.W., X.X., D.L.Z., L.W.D., T.W., H.S.W., C.C. and T.Y. analysed the experimental data and designed the figures. H.W., X.X., C.C., T.Y., D.L.Z., L.W.D. and M.J. co-wrote the manuscript, and all authors contributed to critical discussions of the manuscript.

**Competing interests**

The authors declare no competing financial interests.

| Experiment | Atmosphere | Temperature | NiCl$_2$ Solution concentration | Etching Duration | Length(μm) | Width (nm) | Aspect ratio |
|---|---|---|---|---|---|---|---|
| #1 | Ar:H$_2$=9:1, 150Pa | 1200℃ | 0.01g/L | 60min | 1±0.5 | ~5±3nm | ~200 |
| #2 | Ar:H$_2$=9:1, 150Pa | 1200℃ | 0.01g/L | 180min | ~1.8 | ~11±3nm | ~120 |
| #3 | Ar:H$_2$=9:1, 150Pa | 1200℃ | 0.1g/L | 60min | ~1.5 | ~6±3nm | ~260 |
| #4 | Ar:H$_2$=9:1, 150Pa | 1200℃ | 0.1g/L | 180min | ~2.5 | ~15±3nm | ~160 |
| #5 | Ar:H$_2$=9:1, 150Pa | 1200℃ | 0.5g/L | 10min | <0.01 | N.A. | N.A. |
| #6 | Ar:H$_2$=9:1, 150Pa | 1200℃ | 0.5g/L | 60min | ~2.2 | ~10nm | ~220 |
| #7 | Ar:H$_2$=9:1, 150Pa | 1200℃ | 0.5g/L | 180min | ~3.8 | ~20±3nm | ~190 |
| #8 | Ar:H$_2$=9:1, 150Pa | 1300℃ | 0.01g/L | 60min | ~1.2 | ~50±5nm | ~25 |
| #9 | Ar:H$_2$=9:1, 150Pa | 1300℃ | 0.1g/L | 60min | ~2.3 | ~60±5nm | ~40 |
| #10 | Ar:H$_2$=9:1, 150Pa | 1300℃ | 0.5g/L | 60min | ~3 | ~60±5nm | ~50 |

**Supplementary Table 1 | Summary of nano-particle-assisted-etching experiments under different conditions.** Both the length and width of h-BN trenches exhibit some dependence on experimental parameters, such as pressure of carrying gases, temperature, the concentration of NiCl$_2$ solution, etching duration etc. The width of trenches strongly depends on temperature and duration of etching which is understandable as the etching is based on a thermal activated reaction. However, the evolution of length and the width does not seem to follow the well-known Arrhenius law, most likely due to the involvement of additional etching mechanism, for example, hydrogen-activated etching. The aspect ratios of the trenches are also calculated and insert into the table for comparison.

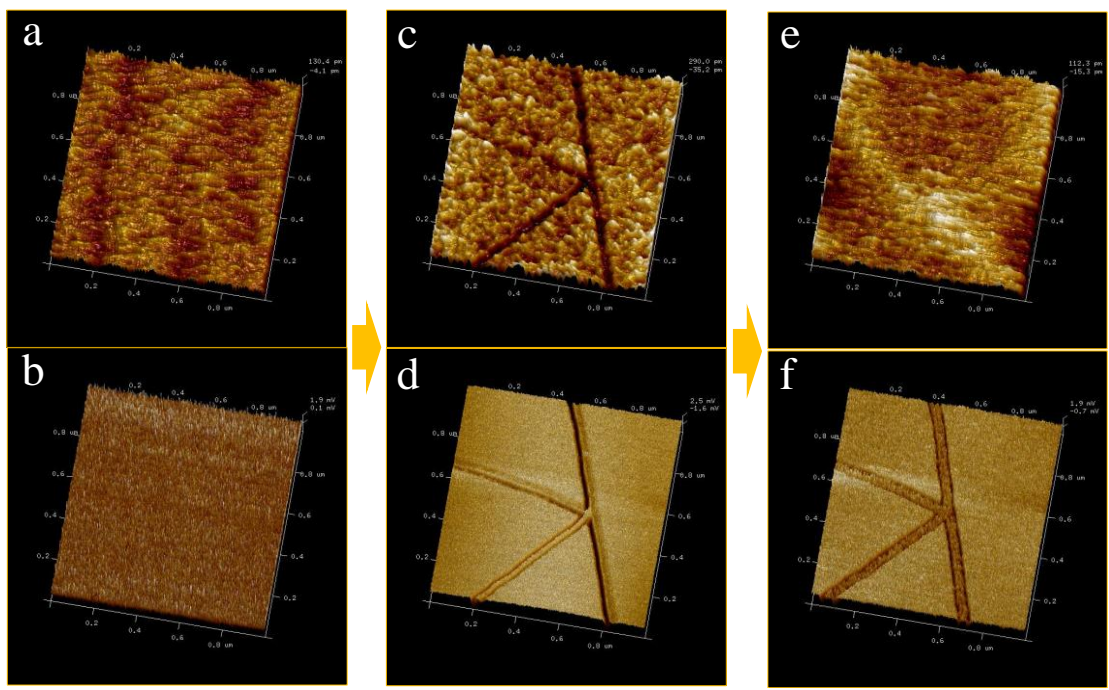

**Supplementary Figure 1 | Fabrication of narrow graphene nanoribbons on the surface of h-BN.** a) 3D AFM height image and b) 3D friction image of h-BN substrate taken after cleaning; c) 3D AFM height image and d) 3D friction image of ~46 nm wide h-BN nano-trenches taken after nano-particle-catalyzed-etching; e) 3D AFM height image and f) 3D friction image after the growth of graphene nano-ribbons in the h-BN nano-trenches.

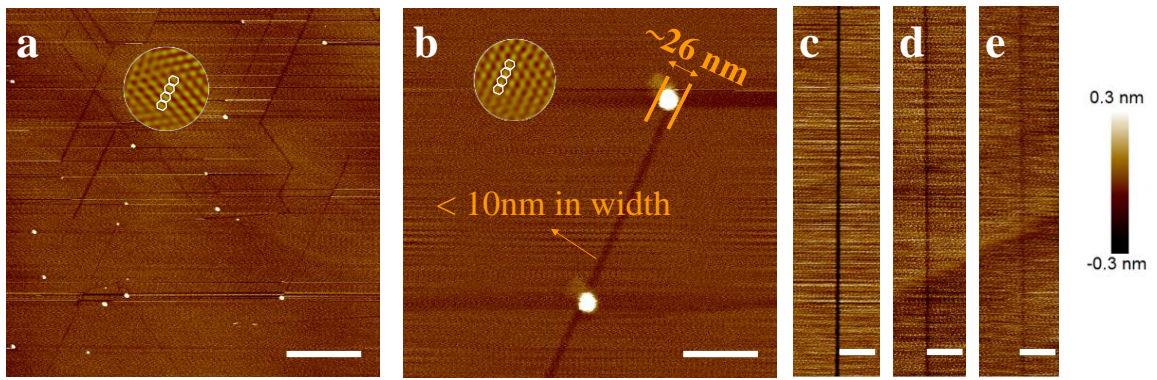

**Supplementary Figure 2 | Images of nano-trenches on the surface of h-BN.** (a-b) AFM height images of h-BN substrate after nano-particle-assisted-etching process, showing the existence of Ni nano-particles and sub-10nm nano-trenches; The circular inset shows atomic-resolution friction image for h-BN, which is Fourier-filtered to improve the signal to noise ratio (SNR). The lattice structures shown in the inset confirm that the crystallographic orientations of the trenches/edges are along zigzag direction. (c-e) Images of narrow nano-trenches in a width of sub-10nm. Scale bar in (a) is 1 μm while scale bars in (b-e) are 100 nm. The height scale for all the AFM height images is 0.6 nm.

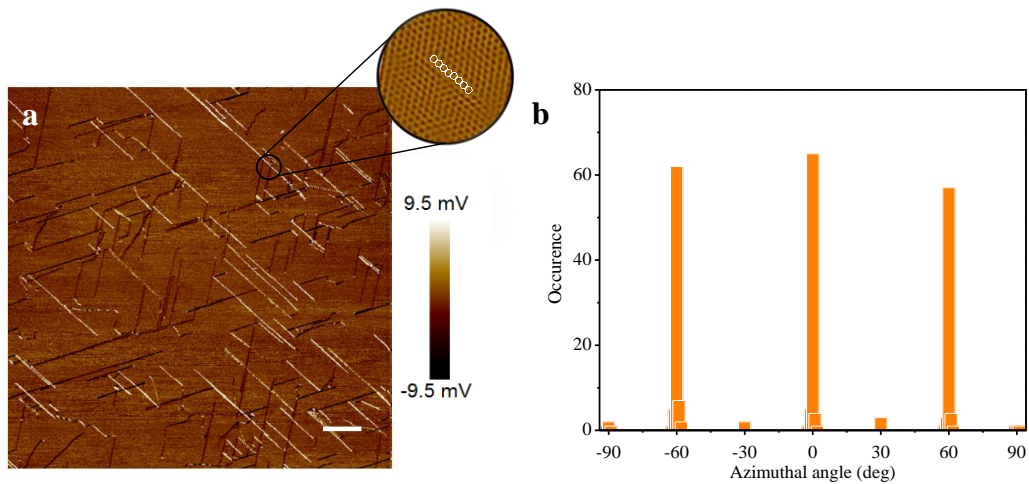

**Supplementary Figure 3 | Orientation distribution of nano-trenches on h-BN.** (a) Scanning probe microscope friction image of nano-trenches on h-BN. Both dark and bright lines indicate individual nanotrenches for the azimuthal-angle measurements. Scale bar: 1 μm. The circular inset shows atomic-resolution friction image for h-BN, which is Fourier-filtered to improve the signal to noise ratio (SNR). The lattice structure shown in the inset confirms that the crystallographic orientations of the trenches/edges are along zigzag direction; (b) Histogram showing the etching-direction distribution of nanotrenches on h-BN. Separation angle of 60° relative to each other indicates an anisotropic etching along equivalent crystallographic directions.

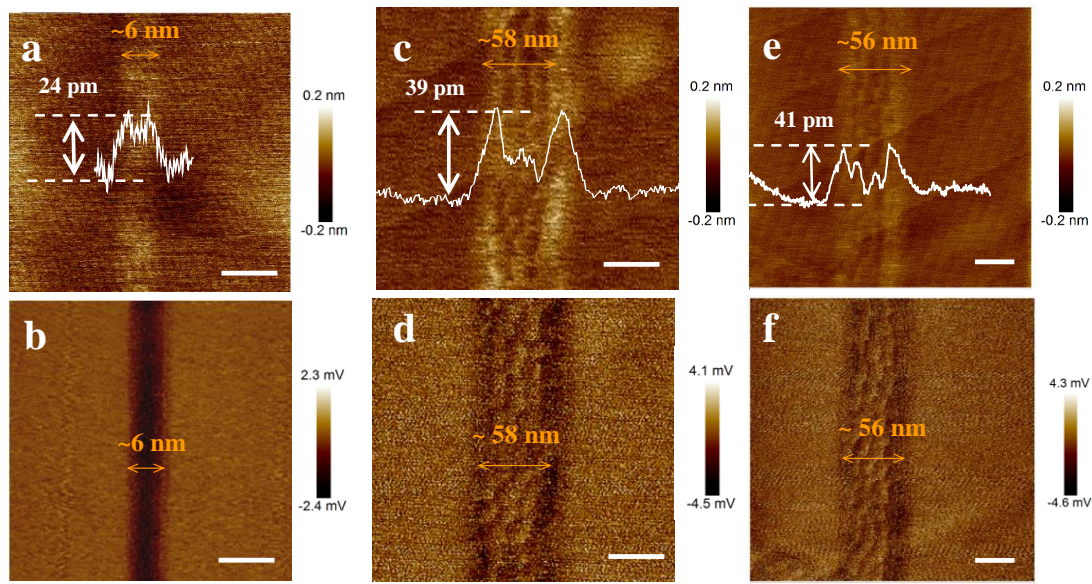

**Supplementary Figure 4 | AFM Images for GNRs on the h-BN.** a) AFM height image and b) friction image of a ~ 6 nm wide GNR embedded in the nano-trenches on the h-BN crystal via template growth, scale bar: 10 nm; c) AFM height image and d) friction image of a ~ 58 nm wide GNR embedded in the nano-trenches on the h-BN crystal via template growth, scale bar: 40 nm; c) AFM height image and d) friction image of a ~ 56 nm wide GNR embedded in the nano-trenches on the h-BN crystal via template growth, scale bar: 40 nm. The height profile of the line-scan across the ribbon indicates that there are obvious out-of-plane deformations along a graphene-BN in-plane boundary. The giant Moiré patterns can be observed on the wide graphene ribbon, and they are found to be stretched along the direction of GNR-h-BN boundary.

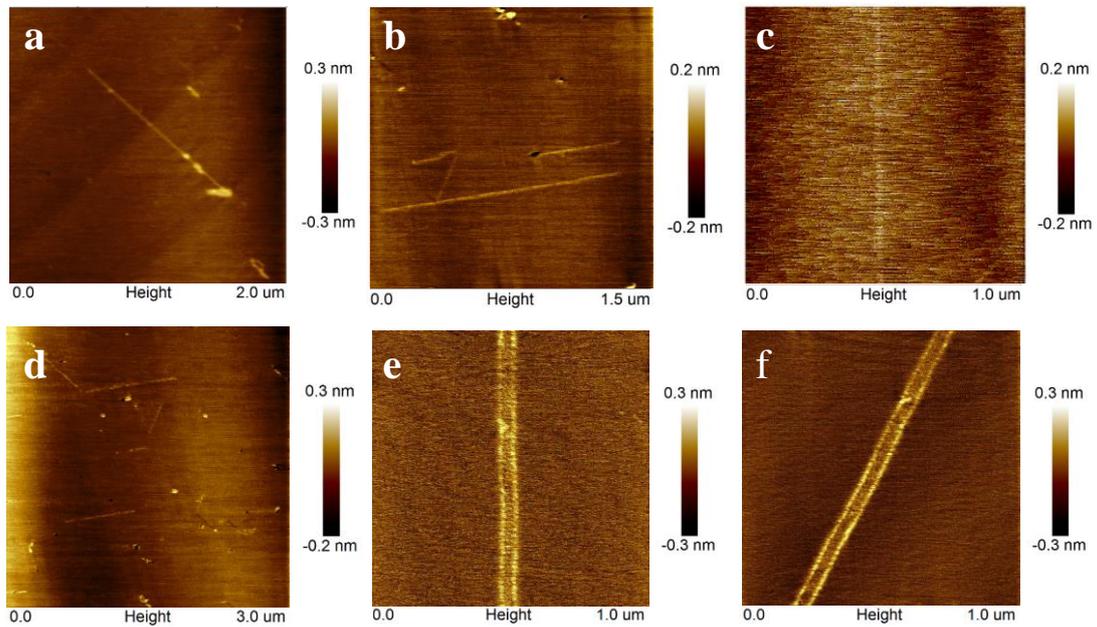

**Supplementary Figure 5 | Images of GNRs embedded in h-BN nano-trench.** (a-d) AFM height images of h-BN substrate after GNR growth process, showing the existence of sub-10nm GNRs; (e-f) AFM height images of about 60 nm-wide GNR on h-BN substrate. The length of GNRs is always more than 1 μm.

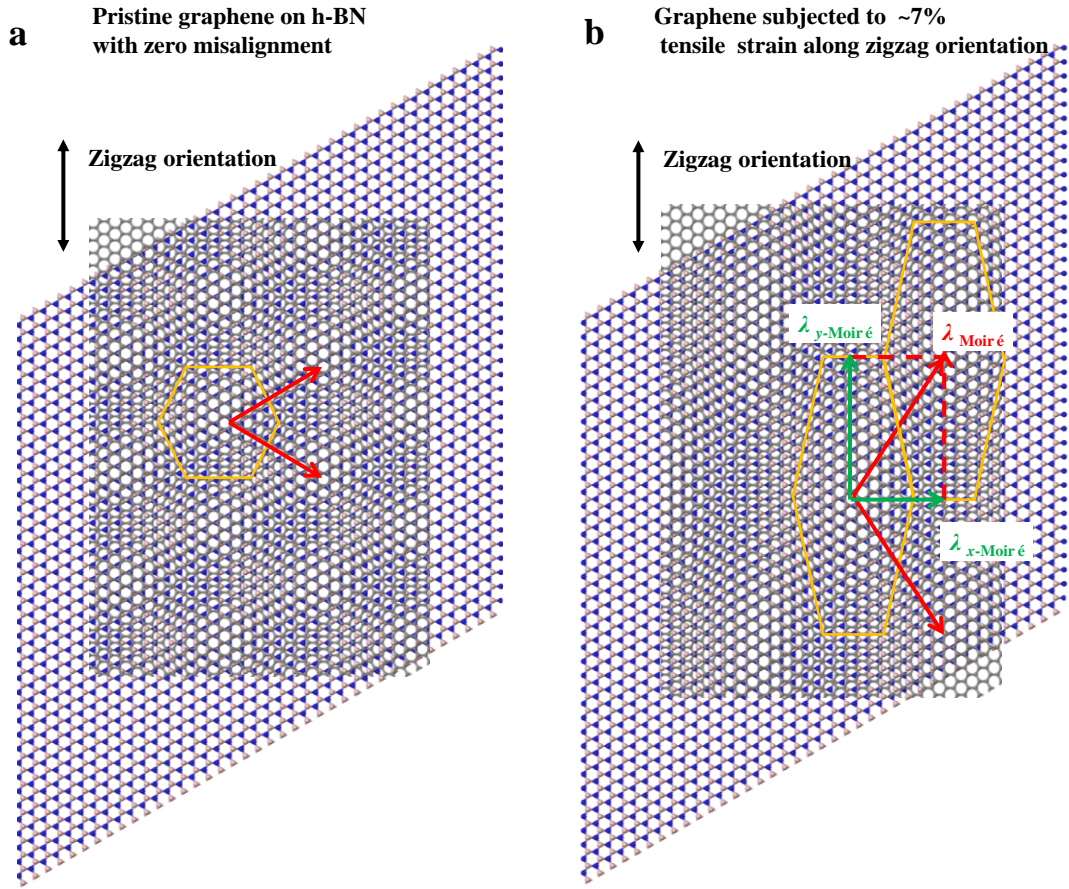

**Supplementary Figure 6 | Illustration of stretched giant Moiré pattern caused by strain in graphene.** (a) Sketch of graphene on h-BN for zero misalignment and an exaggerated lattice mismatch of ~10%. The two superimposed hexagonal lattices cause the emergence of a Moiré pattern. (b) The Moiré pattern is stretched by the tensile strain in graphene along zigzag orientation. $\lambda_{\text{Moiré}}$ represents the lattice vector for stretched Moiré pattern. $\lambda_{y\text{-Moiré}}$ and $\lambda_{x\text{-Moiré}}$ represents the projection of $\lambda_{\text{Moiré}}$ along the zigzag and armchair orientation, respectively.

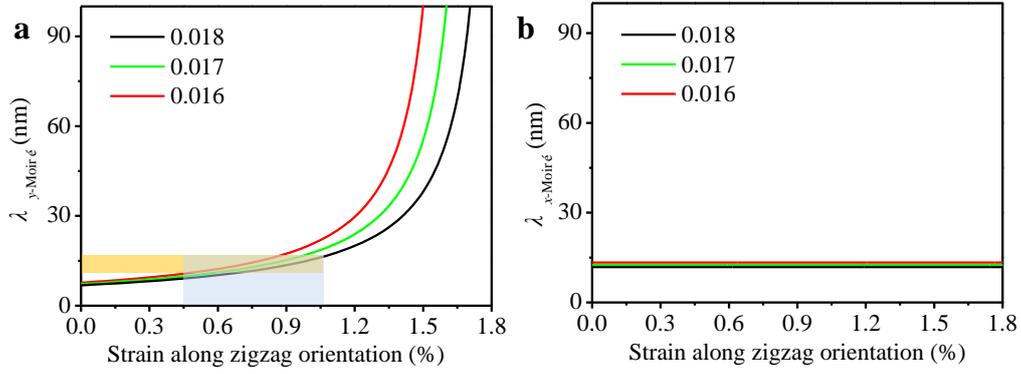

**Supplementary Figure 7 | The wavelength dependence of stretched giant Moiré pattern on strain in graphene precisely aligned with h-BN.** (a) $\lambda_{y\text{-Moiré}}$ *v.s.* strain in graphene along zigzag direction; (b) $\lambda_{x\text{-Moiré}}$ *v.s.* strain in graphene along zigzag direction. Here, cases of the lattice mismatch between graphene and h-BN from 0.016 to 0.018 are simulated.

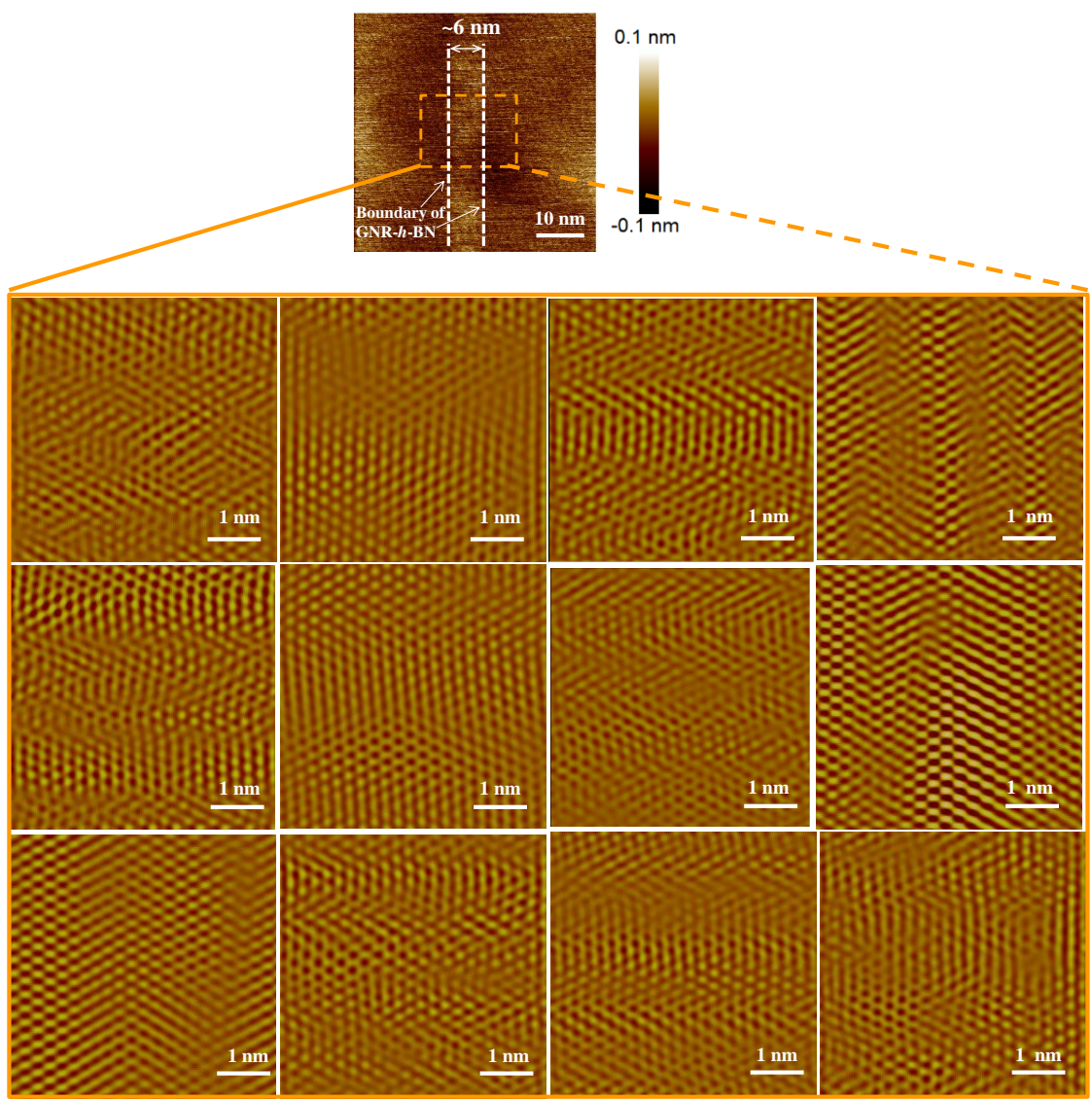

**Supplementary Figure 8 | Atomic-resolution-AFM investigation on graphene "epi-ribbon" embedded in h-BN nano-trench.** The top AFM image shows the area of GNR on h-BN, where the width of GNR is about 6 nm. The atomic resolution scans (5 × 5nm$^2$) below were spatial-continuously taken from the orange box (20 × 15 nm$^2$) in the top AFM image. These atomic images confirm that GNR and the h-BN top-layer form into a uniform monolayer during the CVD growth. By careful examination of atomic resolution AFM, no obvious boundary was found. These AFM images clearly show adjacent graphene nanoribbons grown from opposite step-edge of top h-BN layer have same lattice orientation and coalesce without grain boundary defects. It is also confirmed that the GNR and adjacent h-BN layer have the same lattice orientation and merges at the boundary seamlessly.

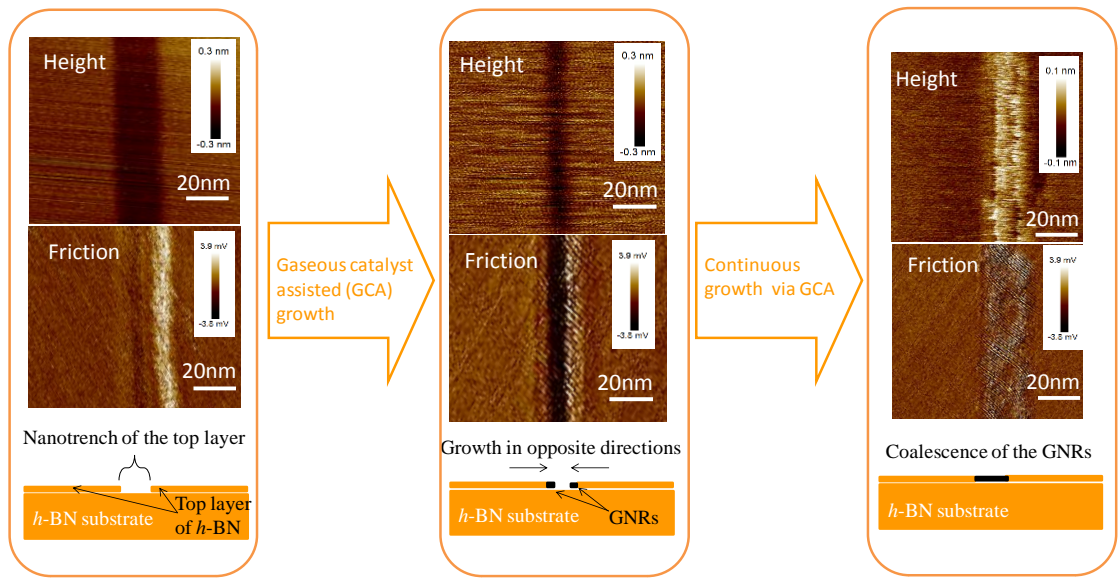

**Supplementary Figure 9 | Formation stages of GNR embedded in h-BN nano-trench.** Schematic, height and friction AFM images for a graphene "epi-ribbon" formed in a nano-trench of h-BN top-layer at different formation stages: (1) after the formation of h-BN nano-trenches; (2) after the beginning of GNRs growth from opposite sides of the nano-trench; (3) after GNRs grown from opposite sides merge together.

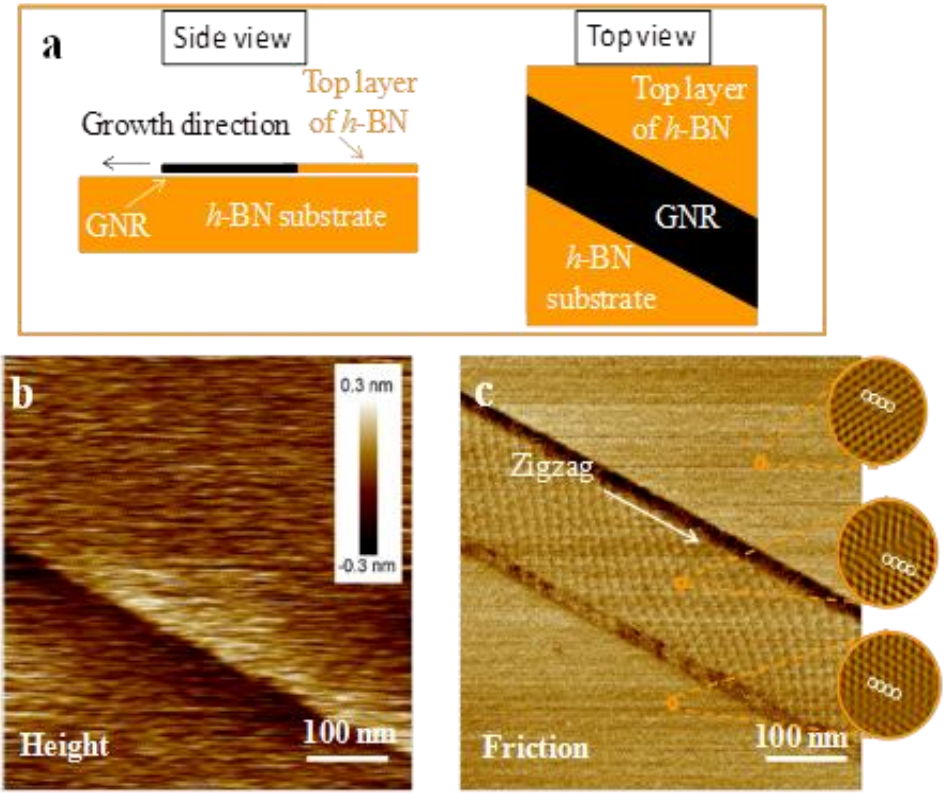

**Supplementary Figure 10 | Graphene "epi-ribbon" along a h-BN step-edge.** (a) Schematics for a graphene "epi-ribbon" formed along a step-edge of h-BN top-layer. Height (b) and friction (c) AFM image of the GNR. The height image reveals that the graphene ribbon seamlessly connects to the h-BN step-edge. In the friction image, the giant Moiré patterns with the periodicity of about 14nm can be clearly seen on the wide GNR. The circular insets in (c) show atomic-resolution friction images (from top to bottom) for the "template" h-BN top-layer, graphene "epi-ribbon" and the "substrate" h-BN layer, respectively. The GNR and h-BN top-layer are found to have the same lattice orientation and merges at the boundary seamlessly.

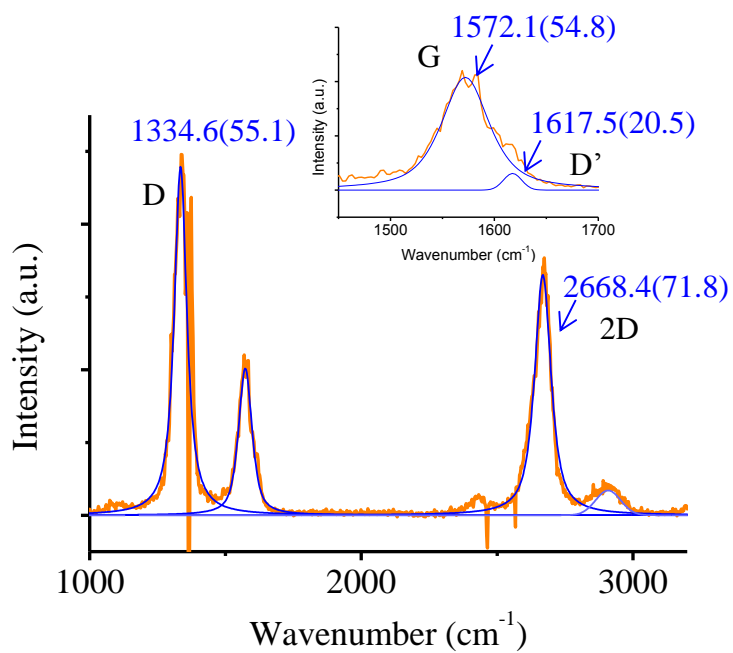

**Supplementary Figure 11 | Raman spectrum of a 15nm-wide graphene nano-ribbon after subtracting the h-BN background signal.** The G-, D- and 2D- bands were fitted with a single Lorentzian line shape. A tiny shoulder (D'-band) appears on the right side of G-band. The full-width at half-maximum (FWHM) for each peak is given in parentheses with the peak location value, and the wavelength of the exciting laser is 488 nm.

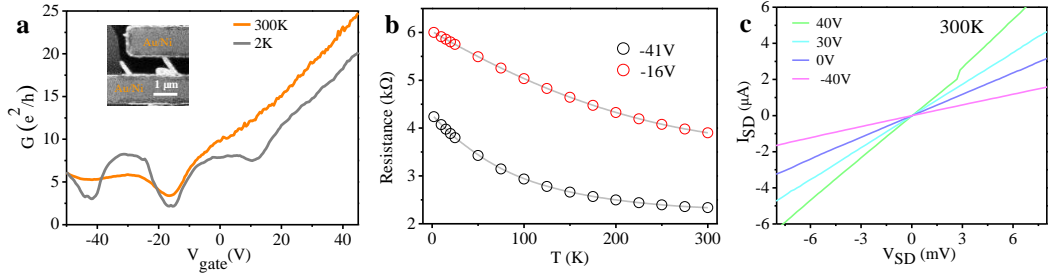

**Supplementary Figure 12 | Electronic transport through a ~58nm GNR device on h-BN.** (a) Conductance (*G*) of a GNR with a width of ~58nm as a function of back gate voltage ($V_{gate}$). In the 58nm-wide GNR, the conductance exhibits three dips at the Dirac point and Secondary Dirac point (SDP), which indicates the existence of giant super-lattice. Inset shows a scanning electron micrograph (SEM) image of the nanoribbon channel corresponding to the transfer curves. The channel length is about 1 μm. Its field effect mobility is about ~3,053 cm$^2$ V$^{-1}$ s$^{-1}$ at 300 K; (b) Resistance under different $V_{gate}$ versus temperature from 2K to 300K for the 58nm-wide GNR, the solid curves are fits based on a simple two-band model, the extracted band gap is about 9 ± 2meV; (c) $I_{SD}$-$V_{SD}$ curves at gate biases $V_{gate}$ ranging from -40 V to 40 V at room temperature.

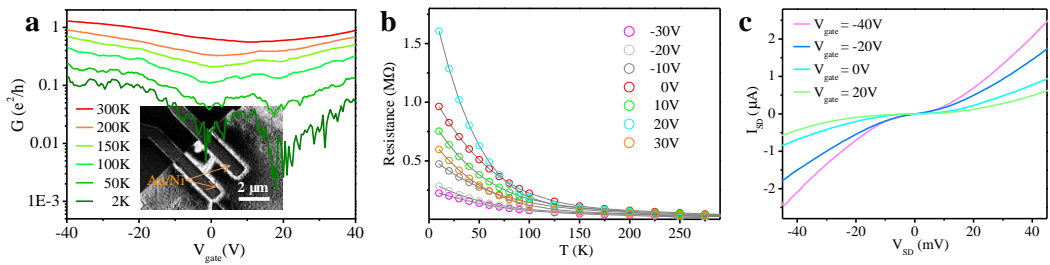

**Supplementary Figure 13 | Electronic transport through a ~20nm GNR device on h-BN.** (a) Conductance (*G*) of a GNR with a width of ~20nm as a function of back gate voltage ($V_{gate}$). Inset shows SEM image of the nanoribbon channel corresponding to the transfer curves. The channel length is about 1.3 μm; Its field effect mobility is about 1,033cm$^2$/Vs cm$^2$ V$^{-1}$ s$^{-1}$ at 300 K; (b) Resistance under different $V_{gate}$ versus temperature from 10K to 300K for the 20nm-wide GNR, the solid curves are fits based on a simple two-band model, the extracted band gap is about 100 ± 11meV; (c) $I_{SD}$-$V_{SD}$ curves for the device in (a) at different gate biases at room temperature.

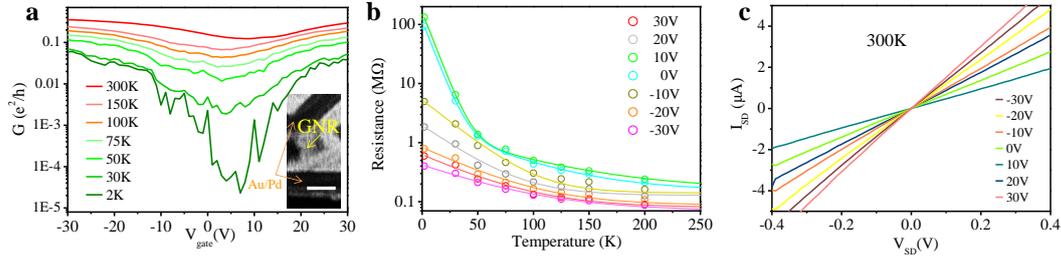

**Supplementary Figure 14 | Electronic transport through a ~15nm GNR device on h-BN.** (a) Conductance ($G$) as a function of back gate voltage ($V_{gate}$). Inset shows SEM image of the nanoribbon channel corresponding to the transfer curves. Scale bar in the inset is 1 μm. The channel length of FET is about 1.5 μm, its field effect mobility is about 916.1cm$^2$/Vs cm$^2$ V$^{-1}$ s$^{-1}$ at 300 K; (b) Resistance under different $V_{gate}$ versus temperature from 25K to 300K for the 20nm-wide GNR, the solid curves are fits based on a simple two-band model, the extracted band gap is about 120 ± 23meV; (c) $I_{SD}$-$V_{SD}$ curves for the device in (a) at different gate biases at room temperature.

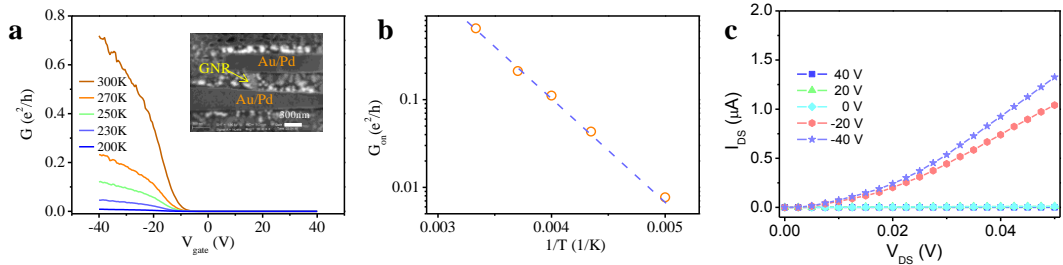

**Supplementary Figure 15 | Electronic transport through a ~5nm GNR device on h-BN.** Conductance (G) of another GNR with an estimated width of about 5 nm. Its conductance can be completely switched-off even at 300K. Inset shows SEM of the nanoribbon channel corresponding to the transfer curves. The channel length is about 300 nm. Scale bar in the inset is 300nm. Its field effect mobility is about 765.8 cm$^2$ V$^{-1}$ s$^{-1}$ at 300 K. The GNR FET in a channel length of about 300nm exhibits rather high $G = 0.7e^2/h$ at room temperature when $V_{gate}$= -40V. Charge carrier was scattered for $(4e^2/h)/(0.7e^2/h) \approx 5$ times. The scattering mean free path (MFP) is estimated to be about 300nm/(5+1)=50nm even without excluding the contribution from the contact; (b) Inset shows the relationship of conductance at $V_{gate}$=-30V v.s. inverse temperature from 200K to 300K, the dashed line is fit to the experimental data according to $G_{on} \sim \exp(\frac{E_g}{2k_B T})$. The extracted band gap is about 489 ± 19meV; (c) $I_{SD}$-$V_{SD}$ curves for the device in (a) at gate biases $V_{gate}$ ranging from -40V to +40V.

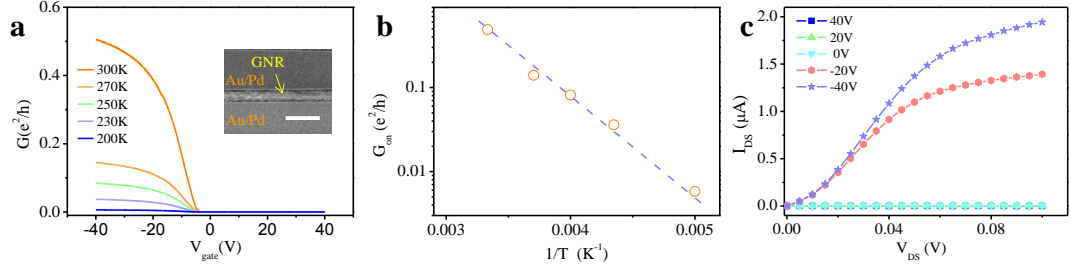

**Supplementary Figure 16 | Electronic transport through a ~5nm GNR device on h-BN.** Conductance (G) of another GNR with an estimated width of about 5 nm. Its conductance can be completely switched-off even at 300K. Inset shows SEM of the nanoribbon channel corresponding to the transfer curves. The channel length is about 500 nm. Scale bar in the inset is 1 μm. Its field effect mobility is about 790.8cm$^2$/Vs cm$^2$ V$^{-1}$ s$^{-1}$ at 300 K; The GNR FET in a channel length of about 500nm exhibits rather high $G = 0.5e^2/h$ at room temperature when $V_{gate}$= -40V. Charge carrier was scattered for $(4e^2/h)/(0.5e^2/h)$≈8 times. The scattering mean free path (MFP) is estimated to be about 500nm/(8+1)≈55.6nm; (b) Inset shows the relationship of conductance at $V_{gate}$=-30V v.s. inverse temperature from 200K to 300K, the dashed line is fit to the experimental data according to $G_{on}$~$\exp(\frac{E_g}{2k_BT})$. The extracted band gap is about 523 ± 37.2meV; (c) $I_{SD}$-$V_{SD}$ curves for the device in (a) at gate biases $V_{gate}$ ranging from -40V to +40V.

# Supplementary Note 1: Simulation of Stretched Moiré Pattern

The schematic in Supplementary Fig. 6 is an explanation to Moiré geometry. When h-BN is overlaid with a graphene layer, new symmetry may emerge leading to the Moiré pattern. Its orientation and wavelength are fixed and can be derived mathematically. The hexagonal lattice of h-BN is defined by vectors ($a_1, a_2$), where $a_1$ and $a_2$ can be written as $a_1 = a(\frac{\sqrt{3}}{2}, -\frac{1}{2})$ and $a_2 = a(\frac{\sqrt{3}}{2}, \frac{1}{2})$, respectively. $a$ represents the lattice constant of h-BN. Similarly, graphene lattice is represented by ($b_1, b_2$). The lattice of precisely-aligned graphene on h-BN substrate is shorter in length by a lattice mismatch factor of δ in the range 1.6% to 1.8%. The factor δ denotes lattice mismatch between h-BN and graphene. $b_1$ and $b_2$ can be written as $b_1 = a(1-\delta)(\frac{\sqrt{3}}{2}, -\frac{1}{2})$ and $a_2 = a(1-\delta)(\frac{\sqrt{3}}{2}, \frac{1}{2})$, respectively. If strain (β) is applied to graphene along zigzag orientation, the $b_1$ and $b_2$ can be written as $b_1 = a(1-\delta)(\frac{\sqrt{3}}{2}(1+\beta), -\frac{1}{2})$ and $a_2 = a(1-\delta)(\frac{\sqrt{3}}{2}(1+\beta), \frac{1}{2})$, respectively. The reciprocal lattice vectors of the Moiré pattern is given by: $k_{moire} = k_{graphene} - k_{h-BN}$. Therefore, the Moiré pattern in real space can be directly derived by numerical method. Here we use 2.50 angstroms as the value of $a$. The projection ($\lambda_{y\text{-Moiré}}$ and $\lambda_{x\text{-Moiré}}$) for the wavelength of Moiré pattern ($\lambda_{Moiré}$) as functions of tensile strain (β from 0 ~ 1.8%) in graphene along zigzag direction are plotted in Supplementary Fig. 7 with lattice mismatch (δ) in the range 1.6% to 1.8%. Here, $\lambda_{y\text{-Moiré}}$ and $\lambda_{x\text{-Moiré}}$ represents the projection of $\lambda_{Moiré}$ along the zigzag and armchair orientation, respectively.

It is found that the value of $\lambda_{y\text{-Moiré}}$ increases with the increase of strain (β) while the value of $\lambda_{y\text{-Moiré}}$ keeps untouched. We noticed that $\lambda_{y\text{-Moiré}}$ has a measured value of 13~16nm in stretched Moiré patterns. (See Fig. 2 and Supplementary Fig. 4). According to the plot in Supplementary Fig. 7, The $\lambda_{y\text{-Moiré}}$ ranging from 13 to 16 nm yields estimated β of 0.45~1.05%. In other words, the GNR aligned with h-BN substrate suffers corresponding tensile strain of 0.75% ±0.3%.

## Supplementary Note 2: Capacitance and Field Effect Mobility

Electronic transport measurements in two-terminal configuration were also carried out to characterize the GNRs grown on h-BN. The gate voltage ($V_g$) dependence of the conductance at different temperature is plotted in Fig. 4 and Supplementary Fig. 12-16. The carrier mobility and on-off ratio of each GNR device can be extracted from the plots.

In a classical regime, the capacitance is completely determined by the object's geometry and a dielectric constant of the medium. If the object's size shrinks to a nanometer scale, a finite DOS which originates from the Pauli exclusion principle should be considered. Low-dimensional systems, having a small DOS, are not able to accumulate enough charge to completely screen the external field. In order to describe the effect of the electric field penetration in a finite-DOS system, quantum capacitance should be taken into account. According to the theoretical calculation (Phy. Rev. B 80, 205402 (2009)), the quantum capacitance can be ignored in devices with thick dielectric layer. As we are using 300nm $SiO_2$ as dielectric layer in these experiments, only classical capacitance is taken into account.

We used three-dimensional electrostatic simulation to calculate $C_{gs}$. "Fast Field Solvers" Software is available at http://www.fastfieldsolvers.com. The simulated structure includes a large highly-doped-Si plane as back-gate, a dielectric bi-layer with the same lateral dimension as back-gate (300nm thickness and $\varepsilon_0$ =3.9, for the bottom $SiO_2$ layer and 10nm thickness and $\varepsilon_0$ =4 for the h-BN top-layer) and a graphene layer with experimental dimension lying ~0.35nm above the dielectric layer and two metal fingers with experimental dimension to represent contacts. ~1 nm grids for GNR are used in simulation. And then, the capacitances are obtained: $C_{gs}$ = 45.1 pF/m for a w=58 nm ribbon (Supplementary Fig. 12), $C_{gs}$ = 5.9 pF/m for a w=20 nm ribbon (Supplementary Fig. 13), $C_{gs}$ = 5.2 pF/m for a w=15 nm ribbon (Supplementary Fig. 14) and $C_{gs}$ = 3.98 pF/m for a w=6 nm ribbon; $C_{gs}$ = 3.71 pF/m for a w=5 nm (Supplementary Fig. 15 and 16). Based on standard transistor model, the intrinsic carrier mobility is $\mu = \frac{\frac{dG}{dV_{gs}} \cdot L}{C_{gs}}$, where $L$ and $w$ represent the channel length and width of GNRs; $G$ represents the channel conductance measured. And then the electrical field mobilities can be derived for the GNR transistors at 300 K (shown in supplementary Fig. 12-16).

# Supplementary Note 3: Simple Two Band (STB) Model

For graphitic materials,[1-4] a simple two band (STB) model are always used to extract the band gap information from the temperature-dependence of resistance measured. Based on this model, the densities of electrons ($n$) and holes ($p$) are given by $n = C_n k_B T \ln(1 + \exp(-\frac{E_C - E_F}{k_B T}))$ and $p = C_p k_B T \ln(1 + \exp(-\frac{E_F - E_V}{k_B T}))$, respectively. Here, $E_F$, $E_C$ and $E_V$ are the energies at the Fermi level, bottom of conduction band and top of valance band, respectively, $k_B$ is the Boltzmann constant and $C_n$, $C_p$ are constants independent of temperature (T). Ignoring the contribution from static scattering centers, the mobility of carriers can be expressed as $\mu_e = A_1 T^{-1}$ or $\mu_h = A_2 T^{-1}$, where $A_1$ and $A_2$ are constants depending on the strength of electron and hole-phonon scattering in graphite. Since the resistivity is given by $\rho = (n\mu_n e + p\mu_p e)^{-1}$, the temperature dependence of resistance can be expressed as:

$$R = \frac{P_1}{\ln(1 + \exp(-\frac{E_C - E_F}{k_B T})) + P_2 \ln(1 + \exp(-\frac{E_F - E_V}{k_B T}))} + R_{contact}$$

Using $E_C - E_F$, $E_F - E_V$, $P_1$, $P_2$ and $R_{contact}$ as the fitting parameters, this model fits well with our experimental data for the most samples at different $V_{gate}$, as shown in supplementary Fig. 12b, 13b and 14b. Based on this model, the energy gaps obtained for the three samples are 9 ± 2meV, 100 ± 11meV and 120 ± 23meV for the GNR samples with a width of 58nm, 20 nm and 15 nm, respectively. The model can rule out the contribution from the contacts (e.g. Schottky barrier). With the decrease of GNR width, the GNR resistance greatly increases. In addition, the current measured decreases greatly and becomes very sensitive to influence from the environment. As such, resistance of GNR cannot be determined accurately. On the other hand, the influence from Schottky barriers becomes pronounced. Therefore, the STB model cannot be applied to the sub-10nm GNR with relatively large $E_g$.

## Supplementary Note 4: Band Gap Estimation in Ultra-narrow GNRs

For the ultra-narrow GNR, it is found that the conductance at On-state exhibits an exponential relationship with inverse temperature from 200K to 300K. It indicates that Schottky barriers (SB) dominate the conductance of the GNR FET because of high work function of Pd and Ni [5,6]. We would like to estimate the band gap ($E_g$) of the narrow GNRs by fitting the conductance near on-state $G_{on} \propto \frac{I_{on}}{I_{off}} \sim \exp(\frac{\phi_{SB}}{k_B T})$ (where $k_B$ is the Boltzmann's constant and $T$ is temperature). Here the height of the Schottky barrier is about half of the gap $E_g$. Thus, $G_{on} \sim \exp(\frac{E_g}{2k_B T})$ [7-9] The extracted values of band gaps for the 5-6 nm GNRs are 489.4 ±19.0meV, 523 ± 37.2meV, 468 ± 7.2meV and 476 ±20meV, respectively.

# Supplementary References

1 Novoselov, K.S. *et al.*, Electric field effect in atomically thin carbon films. *Science* **306**, 666-669 (2004).

2 E. Dujardin, *et al*., Fabrication of mesoscopic devices from graphite microdisks. *Appl. Phys. Lett.* **79**, 2474-2476 (2001).

3 Wang, H., Choong, C., Zhang, J., Teo, K. L. and Wu, Y. Differential conductance fluctuation of curved nanographite sheets in the mesoscopic regime. *Solid State Communications* **145**, 341-345 (2008).

4 C. Berger, *et al*., Electronic confinement and coherence in patterned epitaxial graphene. *Science* **312**, 1191-1196 (2006).

5 Javey, A., Guo, J., Wang, Q., Lundstrom, M. & Dai, H. Ballistic carbon nanotube field-effect transistors. *Nature* **424**, 654-657 (2003).

6 Li, X., Wang, X., Zhang, L., Lee, S. & Dai, H. Chemically derived, ultrasmooth graphene nanoribbon semiconductors. *Science* **319**, 1229-1232 (2008).

7 Chen, Z., Lin, Y.-M., Rooks, M. J., Avouris, P. *Physica E: Low-dimensional Systems and Nanostructures*, **40**, 228-232 (2007).

8 Xia, F., Farmer, D., Lin, Y., and Avouris, P. Graphene field-effect-transistors with high on/off current ratio and large transport band gap at room temperature. *Nano Letters* **10**, 715-718 (2010).

9 Hwang, W. S. *et al*., Graphene nanoribbon field-effect transistors on wafer-scale epitaxial graphene on SiC substrates. *APL Mater.* **3**, 011101 (2015).